\documentclass[12pt]{JHEP3} % 10pt is ignored!
\input colordvi

\usepackage{epsfig,multicol}

\newcommand{\lsim}   {\mathrel{\mathop{\kern 0pt \rlap
  {\raise.2ex\hbox{$<$}}}
  \lower.9ex\hbox{\kern-.190em $\sim$}}}
\newcommand{\gsim}   {\mathrel{\mathop{\kern 0pt \rlap
  {\raise.2ex\hbox{$>$}}}
  \lower.9ex\hbox{\kern-.190em $\sim$}}}
\def\be{\begin{equation}}
\def\ee{\end{equation}}
\def\ba{\begin{eqnarray}}
\def\ea{\end{eqnarray}}

\def\nuebar{\bar{\nu}_e}
\def\numubar{\bar{\nu}_{\mu}}

\def\nue{\nu_e}
\def\numu{\nu_\mu}

\def\Dm2{\Delta m^2}
\def\s2t{\sin^2{2\Theta}}
\def\ap{\approx}

\def\Ue{|U_{e3}|}
\def\UeUe{|U_{e3}|^2}
\newcommand\fverb{\setbox\pippobox=\hbox\bgroup\verb}
\newcommand\fverbdo{\egroup\medskip\noindent%
                        \fbox{\unhbox\pippobox}\ }
\newcommand\fverbit{\egroup\item[\fbox{\unhbox\pippobox}]}
\newbox\pippobox
\title{\textBlack {
The HLMA project: determination of high $\Dm2$ LMA mixing parameters
and constraint on $\Ue$ with a new reactor neutrino experiment \\}
\textBlack}

\author{Stefan Sch\"onert\thanks{Corresponding author.}\\
        Max-Planck-Institut f\"ur Kernphysik, Saupfercheckweg 1, 
        D-69117 Heidelberg, Germany\\
        E-mail: \email{stefan.sch\"onert@mpi-hd.mpg.de}}

\author{Thierry Lasserre\\
        Max-Planck-Institut f\"ur Kernphysik, Saupfercheckweg 1, 
        D-69117 Heidelberg, Germany\\
        DSM/DAPNIA/SPP, CEA/Saclay,
        91191 Gif-Sur-Yvette CEDEX, France\\
        E-mail: \email{thierry.lasserre@mpi-hd.mpg.de}}

\author{Lothar Oberauer\\
        INFN, Laboratori Nazionali del Gran Sasso, I-67010 Assergi (AQ), Italy\\
        E-mail: \email{lothar.oberauer@lngs.infn.it}}

\abstract{
In the forthcoming months, the KamLAND experiment will probe the 
parameter space of the solar large mixing angle (LMA) MSW solution 
as the origin of the solar neutrino 
deficit  with $\nuebar$'s from distant nuclear reactors.
If however the solution realized in nature is such that $\Dm2_{sol} \gsim 2 \cdot 10^{-4}$~eV$^2$ (thereafter named the HLMA region), KamLAND
will only observe a rate suppression but no spectral distortion and hence it
will not have the optimal sensitivity to measure the mixing parameters. In this case, we propose a new medium baseline reactor experiment located at Heilbronn (Germany) to pin down the precise value of the solar mixing parameters. In this paper, we present the Heilbronn detector site, we calculate the $\nuebar$ interaction rate and the positron spectrum expected from the surrounding nuclear power plants. We also discuss the sensitivity of such an experiment to $\Ue$ in both normal and inverted neutrino mass hierarchy scenarios. We then outline the detector design, estimate background signals induced by natural radioactivity as well as by in-situ cosmic ray muon interaction, and discuss a strategy to detect the anti-neutrino signal 'free of background'. 
}


\begin{document} 

%\maketitle  IS IGNORED %%%%%%%%%%%

\section{Introduction}

Over the last years, 
measurements of atmospheric and solar neutrinos gave increasing evidence 
for massive neutrinos and lepton mixing. Recent results of the 
SNO experiment combined with those of Super-Kamiokande showed for the first
time directly that neutrino flavour conversion occurs \cite{SNO,SKsol}.

Whilst atmospheric neutrino measurements 
(dominant channel: $\nu_{\mu} \to \nu_{\tau}$)  
confine the oscillation parameters to
$1.4 \times 10^{-3} < \Delta m^2_{atm}  < 4.2 \times 10^{-3}$~eV$^2$ 
and  $\sin^2 2\Theta_{atm} > 0.9 $ (90\%~CL) \cite{SKatm_Messier}, 
solar neutrino data (dominant channel: $\nu_{e} \to \nu_{\mu,\tau}$)
allow various disjoint areas in oscillation parameter space ranging from 
$10^{-10}$ to $10^{-3}$ eV$^2$ \cite{FitBahcall,FitFogli,FitGiunti}. 
They are commonly
referred to as small mixing angle  (SMA), large mixing  (LMA), 
LOW and (quasi-) vacuum (VAC) solution.
The main objective of ongoing and upcoming solar and reactor neutrino
experiments is to unambiguously identify the solution and 
determine with high accuracy $\Delta m^2_{sol} $ and the respective mixing angle.

Solutions with $\Dm2_{sol} < 10^{-6}$ eV$^2$ {\em or} small mixing 
angles can be probed best with solar neutrino experiments at 
sub-MeV energies \cite{StefanTaup}. Values of $\Dm2_{sol} > 10^{-6}$ eV$^2$ 
{\em and} large mixing angles can be studied both with long-baseline nuclear 
reactor neutrino experiments \cite{KamPropUS,KamLAND,BorexReactor} and with sub-MeV 
solar neutrinos \cite{bx,lens,xmass}. 
Large mixing, in particular at large $\Dm2_{sol}$ values,
is favoured when combining all solar neutrino data. 
The LMA solution gives the best $\chi^2$ values in global analysis
for $\Dm2_{sol} = (3.7-6.3)\times 10^{-5}$~eV$^2$ 
and  $\tan^2 \Theta_{sol} \simeq 0.35-0.38$ 
($\sin^2 2\Theta_{sol} \simeq 0.77 - 0.80$) \cite{FitBahcall,FitFogli,
FitGiunti,FitSmy}. 
Large part of the $10^{-4}$~eV$^2$ range is allowed by the current data. 
It is limited by the CHOOZ reactor experiment at 
$\sim 1 \times 10^{-3}$~eV$^2$ \cite{CHOOZ2nu}.
In the following we shall focus on the potential of long-baseline
reactor neutrino experiments to observe neutrino oscillations 
and to determine $\Dm2_{sol}$ in case that the LMA solution is realized 
in nature.

Electron anti--neutrinos ($\nuebar$) from nuclear reactors
have a continous energy spectrum up to about 10~MeV. 
They can be detected via inverse beta decay on protons 
$\nuebar p \to e^+ n$ for $E_{\nuebar} > E_{thr} \sim  1.8$~MeV, and their
energy is derived from the measured positron kinetic energy as
$E_{\nuebar} \simeq  E_{e^+} + E_{thr}$.
The inverse beta decay cross section, including recoil, weak magnetism,
and radiative corrections, can be found in \cite{VogelBeacom}.

If the distance  between nuclear reactor and detector 
is larger than, or equal to the oscillation length, 
neutrino oscillations become observable as an integral 
reduction of the interaction rate, 
as well as a periodic modulation of the continous 
$\nuebar$ spectrum. For two-neutrino mixing the 
survival probability is 
\be
\label{2nuSP}
P_{\nuebar \to \nuebar} =
1-\sin^2 2\Theta_{sol} \cdot \sin^2 \left( 1.27\, \frac{\Dm2_{sol} [{\rm eV}^2]\, 
L[{\rm m}]} {E_{\nuebar} [{\rm MeV}]}\right) ,
\ee
with the mixing angle $\Theta_{sol}$, the mass difference $\Dm2_{sol}$,
the reactor--detector distance $L$ and the neutrino energy $E_{\nuebar}$. 
Since the mixing angle is expected to be large,
the periodic modulation of the energy spectrum 
should become clearly visible and the value of $\Dm2_{sol}$
could be derived with high accuracy. 
 
If however, for a given value of $\Dm2_{sol}$ the baseline is choosen
too long, adjacent peaks cannot be resolved 
experimentally. The shape of the positron spectrum then
appears unchanged, whilst its normalization is determined
by the mixing angle only, independent of the actual value of $\Dm2_{sol}$.  
In this case only lower limits on $\Dm2_{sol}$ could be derived. 
An additional smearing can arise, if several reactors with different
baseline distances contribute to the interaction rate.

Currently, there are two experiments which have the sensitivity
to explore the parameter space of the solar LMA solution with
$\nuebar$'s from nuclear reactors:
the neutrino signal in the KamLAND experiment in Japan 
is dominated by nuclear power reactors at a distance of 160~km. 
It is thus sensitive to probe distortions of the positron spectrum
for values of $\Dm2_{sol} \sim 2\times 10^{-5}$ to 
$\sim 2\times 10^{-4}$~eV$^2$ \cite{Barger,Barbieri,GouveaPena}.
The BOREXINO experiment in Italy has a  characteristic 
reactor baseline distance of about 750~km, thus would observe
spectral distortions  for 
$\Dm2_{sol} \sim  4\times 10^{-6}$ to $\sim 4\times 10^{-5}$ eV$^2$.
Both upper limits are approximate only and depend on the energy resolution 
that will be achieved in the experiments. A resolution of  
$\lsim$~10\% ($\sigma$) at 1~MeV is expected in KamLAND  \cite{KamPropUS}
and for BOREXINO an even better resolution is expected as the design
goal for the light yield is about 400 pe/MeV  \cite{bx}.

%%%%%%%%%%%%%%%%%%%%%%%%%%%%%%%%%%%%%%%%%%%%%%%%%%%%%%%
%%%%% changes for version 3 required by the referee%%%%
Several authors 
%\cite{Barbieri,StefanTaup,StrumiaVissani,PetcovNHIH} 
pointed out that a dedicated reactor neutrino oscillation experiment is 
needed, if $\Dm2_{sol}~\gsim~2\times~10^{-4}$~eV$^2$. 
The authors of Ref. \cite{Barbieri} concluded that a baseline shorter than 
that of KamLAND is required in order to determine $\Dm2_{sol}$ in the above
quoted range. 
The authour of Ref. \cite{StefanTaup} discussed a reactor experiment with
baseline of {\it a few 10 km} to investigate the high $\Dm2$ range 
of the LMA parameter space. First experimental details of this study
were presented in \cite{StefanNOON01}.
The authors of Ref. \cite{StrumiaVissani} studied the achievable accuracy of 
$\Dm2_{sol}$ and $\rm{ tan^2 \Theta}$ in a hypothetical reactor experiment
with a baseline of 20 km and 3000 events/year. Finally, the authors 
of Ref. \cite{PetcovNHIH} discussed the positron 
energy spectra for a generic reactor experiment with 20~km 
baseline and the accuracy of $\Dm2_{sol}$ determination. 
Moreover, the authors pointed out that for a distinct 
combination of mixing parameters,
one can distinguish normal from inverted mass hirarchy. 
%%%%%%%%%%%%%%%%%%%%%%%%%%%%
% text of version 2:
%It has been pointed out in Refs.
%\cite{Barbieri,StefanTaup,StrumiaVissani,PetcovNHIH} that a dedicated 
%reactor neutrino oscillation experiment is needed, if
%$\Dm2_{sol}~\gsim~2\times~10^{-4}$~eV$^2$.
%%%%%%%%%%%%%%%%%%%%%%%%%%%%
In summary, a new reactor neutrino experiment would become 
mandatory in the case that KamLAND observes a reduction of the 
event rate but could not resolve the characteristic oscillation 
pattern of the positron energy spectrum.
  
In this paper we propose a location and a design 
for a reactor neutrino experiment which would be dedicated to study, 
with high sensitivity, neutrino oscillations in  the HLMA parameter range, 
i.e. $2\times~10^{-4}~\lsim~\Dm2_{sol}~\lsim 10^{-3}$~eV$^2$. 
Moreover, we discuss a new approach to constrain $\Ue$ 
from the observation of ``atmospheric'' driven oscillation 
imprinted on the positron spectrum. 
Finally, we estimate for the proposed location and detector design, the 
background signals induced by natural radioactivity as well as by 
in-situ cosmic ray muon interaction, and discuss a strategy
to detect the neutrino signal 'free of background'.

\section{Experimental Site}

Requiring the first oscillation peak to appear at about 1~MeV above threshold 
(i.e $E_{\nuebar}\simeq2.8$~MeV), the minimal distance between reactor and 
detector for $\Dm2_{sol} = 2\times 10^{-4}$~eV$^2$ given by Eq.~(\ref{2nuSP}) 
should be approximately 17~km. The detector should 
be located sufficiently deep underground to reduce the cosmic ray 
muon flux. Furthermore, the $\nuebar$-interaction rate 
should be dominated by a single baseline only
to minimize the smearing of the oscillation pattern from several sources. 

Various location have been investigated over the last year. 
Our most favoured site is a saltmine close 
to Heilbronn, in the south-west of Germany. More than 2000 cavernes with 
approximate dimensions of 15~m~$\times$~15~m~$\times$~150~m 
have been excavated at a depth between 180 and 240 m (480 to 640 meter
of water equivalent (mwe)). Two reactors 
are in the close, nearly equidistant vicinity: 
Neckarwestheim (2 cores, 6.4 GW$_{th}$) in southern and Obrigheim 
(1 core, 1.1 GW$_{th}$) in northern direction. Since the saltmine 
is extended about six kilometers in north-south direction,
it is possible to select a location with a common baseline of
exact 19.5~km to each of the two reactors. 
%Further reactors which contribute to the signal rate are listed in Tab.~\ref{tab:reactorData}. 
It is noteworthy that the Obrigheim reactor might be shut down in the very near future. 
In this case the baseline could be selected between 14 and 20~km,
depending to which parameters the experiment needs to be optimized to.
In this paper we take as generic location the position with 
equidistant baselines of 19.5~km to the reactors Neckarwestheim and Obrigheim.

\section{Anti-neutrino interaction rate}

The $\nuebar$ spectrum above detection threshold is the result of $\beta^-$ decays of $^{235,238}$U and $^{239,241}$Pu fission products. Mesurements for $^{235}$U and $^{239,241}$Pu and theoretical calculations for $^{238}$U are used to evaluate the $\nuebar$ spectrum \cite{Schreckenbach:1985ep,Hahn:1989zr}; its overall normalisation is known to about $1.4\%$ \cite{Declais} and its shape to about $2\%$ \cite{Bugey}. As a nuclear reactor operates, the fission element proportions evolve in time; as an approximation we use an averaged fuel composition typical during a reactor cycle corresponding to  $^{235}$U (55.6 \%), $^{239}$Pu (32.6 \%), $^{238}$U  (7.1 \%) and $^{241}$Pu (4.7 \%). The mean energy release $(W)$ per fission is then 203.87 MeV and the energy weighted cross section for $\nuebar p \rightarrow n e^+$ amounts to $<\sigma >_{\rm fission} = 5.825 \times 10^{-43}$~cm$^2$ per fission. The reactor power ($P_{th}$) is related to the number of fissions per second ($N_f$) by $N_f = 6.241 \times 10^{18} {\rm sec}^{-1} \cdot (P_{th}[{\rm MW}]) / (W[{\rm MeV}])$. The event rate ($R_L$) at a distance $L$ from the source, assuming no--oscillation, is then $R_L~=~N_f~\cdot~<~\sigma~>_{\rm fission}\cdot n_p\cdot~1/(4\pi L^2) $, where $n_p$ is the number of protons of the target. For the purpose of simple scaling, a reactor with a power of 1~GW$_{th}$ induces a rate of 447.8 events per year in a detector with $10^{31}$~protons at a distance of 10~km. All relevant European nuclear plants are added in turn in order to compute the $\nuebar$ interaction rate  at the Heilbronn site. Five of them (see Tab.~\ref{tab:reactorData}) contribute to $\sim 92\%$ of the total rate $R_0$; the other reactors contribute less than $1\%$ to the total flux each, and 8 \% in total. Assuming  all reactors running at their nominal power, one expects $\sim 1150$ $\nuebar$ interactions per year in a target containing $10^{31}$ protons (as an example, a PXE-based scintillator with a mass of 194 tons contains $10^{31}$ protons).   
\TABLE[h]{
\centering
\caption{{\bf Nuclear reactor data and interaction rate at the Heilbronn site.}
The neutrino interaction rates $R_L, R_0$ are given for a target containing $10^{31}$ protons with the nuclear power plant running at 100\% of their nominal power (typical mean values of the power vary between 80\% and 90\%). A PXE-based scintillator (C$_{16}$H$_{18}$, molar weight=210 g/mol)  with a mass of 194 tons contains $10^{31}$ protons.}
\label{tab:reactorData}
\begin{tabular}{lccccc}
\noalign{\bigskip}
\hline
Reactor & Distance [km] & Power [GW$_{\rm th}$] & $R_L$ & $R_L/R_{0}$ \\
\hline
Neckarwestheim & $19.5$ & $6.388$   & $754$ & $66\%$  \\
Obrigheim      & $19.5$ & $1.057$   & $125$ & $11\%$  \\
Philipsburg    & $54 $  & $6.842$   & $107$ & $9\%$   \\
Biblis         & $80 $  & $7.420$   & $52 $ & $4\%$   \\
Grundremmingen & $117$  & $7.986$   & $26 $ & $2\%$   \\ 
Others (Europe)& $>100$   & $\sim 293$ & $86$ & $8\%$ \\
\noalign{\smallskip}
\hline
\end{tabular}
}

\section{Neutrino oscillation signatures}
A three-neutrino mixing scenario is required to explain both atmospheric and solar anomalies. In that case, flavour eigenstates $\alpha = e,\mu,\tau$ and mass eigenstates $i=1,2,3$ are related through the Pontecorvo-Maki-Nakagawa-Sakata (PMNS) mixing matrix $U$ via the relation $\nu_{\alpha}=\mathop{ \sum_{i=1}^{3}{U_{\alpha i}}} \nu_{i}$ \cite{Pontecorvo,MSN}. Assuming a ``normal'' mass hierarchy scenario, $m_{1} < m_{2} < m_{3}$, the $\nuebar$ survival probability can be written \cite{CHOOZU13,PetcovNHIH}
%\footnote{For these values of distance and energy, Earth matter effects can be neglected.}:
%
\begin{eqnarray}
\label{3nuSP}
P_{\nuebar \to \nuebar} & =  & 1 - 2 \UeUe (1-\UeUe) \left( 1 - \cos{\frac{\Dm2_{31} L}{2 E}} \right)  \\
&   & - \frac{1}{2}  (1-\UeUe)^2 \sin^2{2\Theta_{12}} \left( 1 - \cos{\frac{\Dm2_{21} L}{2 E}} \right)\nonumber  \\
&   & + 2  \UeUe (1-\UeUe)  \sin^2{\Theta_{12}} \left( \cos{\left( \frac{\Dm2_{31} L}{2 E} - \frac{\Dm2_{21} L}{2 E} \right)} - \cos{\frac{\Dm2_{31} L}{2 E}} \right) \nonumber.
\end{eqnarray}
The first two terms of Eq.~\ref{3nuSP} contain respectively the atmospheric driven ($\Dm2_{31} = \Dm2_{atm}$) and solar driven ($\Dm2_{21} = \Dm2_{sol}$, $\Theta_{12} \sim \Theta_{sol}$) contributions, while the third term, absent from {\it any} two-neutrino mixing model, is an interference between solar and atmospheric driven oscillations. We only notice here that $U_{e3}$ is the PMNS matrix element that couples the heaviest neutrino field to the electron field. 
The relative neutrino interaction rate and spectrum compared with the no-oscillation case is given, for a single reactor at a distance L from the detector, by $R/R_{L} = \int{ \Phi (E) \cdot P_{ \nuebar \rightarrow \nuebar } (E) \cdot dE } $, where $\Phi$ is the $\nuebar$ production energy spectrum weighted by the inverse $\beta-$decay cross section. In what follows, the positron energy spectra expected from the five power plants listed in Tab.~\ref{tab:reactorData} are added in turn to obtain the signal at the Heibronn site. If the KamLAND experiment points out any part of the HLMA region as the correct solution of the solar neutrino problem, the experimental signature at the Heilbronn site will be a reduction of the $\nuebar$ rate as well as a distortion of the positron energy spectrum. 
\subsection{Two-neutrino mixing analysis}
If $\Ue$ vanishes the lepton mixing is radically simplified and the relation \ref{3nuSP} reduces then to \ref{2nuSP}, with $\Dm2_{21}=\Dm2_{sol}$ and $\Theta_{12}=\Theta_{sol}$; this approximation has been used by previous reactor neutrino experiments \cite{CHOOZ2nu,PV}. If $\UeUe < 0.01$ the two-neutrino model remains valid with the small correction $\sin^2{2\Theta_{sol}}=(1-\UeUe)^2 \sin^2{2\Theta_{12}}$. The sensitivity of the experiment depends obviously on the background level that can be evaluated from the time variation of the reactor signal; at this stage, assuming no background and nuclear power plants running at their full capacity, a detector containing $10^{31}$ target protons would be sensitive to a $10\%$ suppression in rate at a 3~$\sigma$ level after one year of data taking.  Fig.~\ref{fig:ratecontour} displays the contour lines of equal rate suppression, and thus the potential sensitivity at the Heilbronn site over the LMA region satisfying $\Dm2_{sol} \gsim 3 \times 10^{-5}$ eV$^2$. In what follows, the finite energy resolution of the detector has not been included. The expected positron energy spectrum is displayed in Fig.~\ref{fig:posspectrum} for several combinations  $\Dm2_{sol} - \s2t_{sol}$. Since $\nuebar$'s with different energies arrive at the Heilbronn detector with different phases, shape distortions of the positron energy spectrum are expected. These deformations depend strongly on the $\Dm2_{sol}$ value in the range $1~-~10~\times~10^{-4}$~eV$^2$; this constitutes the main advantage of a medium baseline reactor experiment over other choices, in order to measure accurately high $\Dm2_{sol}$ values lying in the HLMA area. The sensitity to a distorted spectrum could extend up to a few times $10^{-3}$ eV$^2$, depending on the achievable energy resolution. This overlaps significantly with the parameter space probed by the CHOOZ and Palo-Verde experiments \cite{CHOOZ2nu,PV}. Whereas the precision measurement of $\Dm2_{sol}$ depends mainly on the energy resolution, the accuracy of the determination of the solar mixing angle will rely on the statistics as well as the control of the various backgrounds.
\EPSFIGURE[htb]{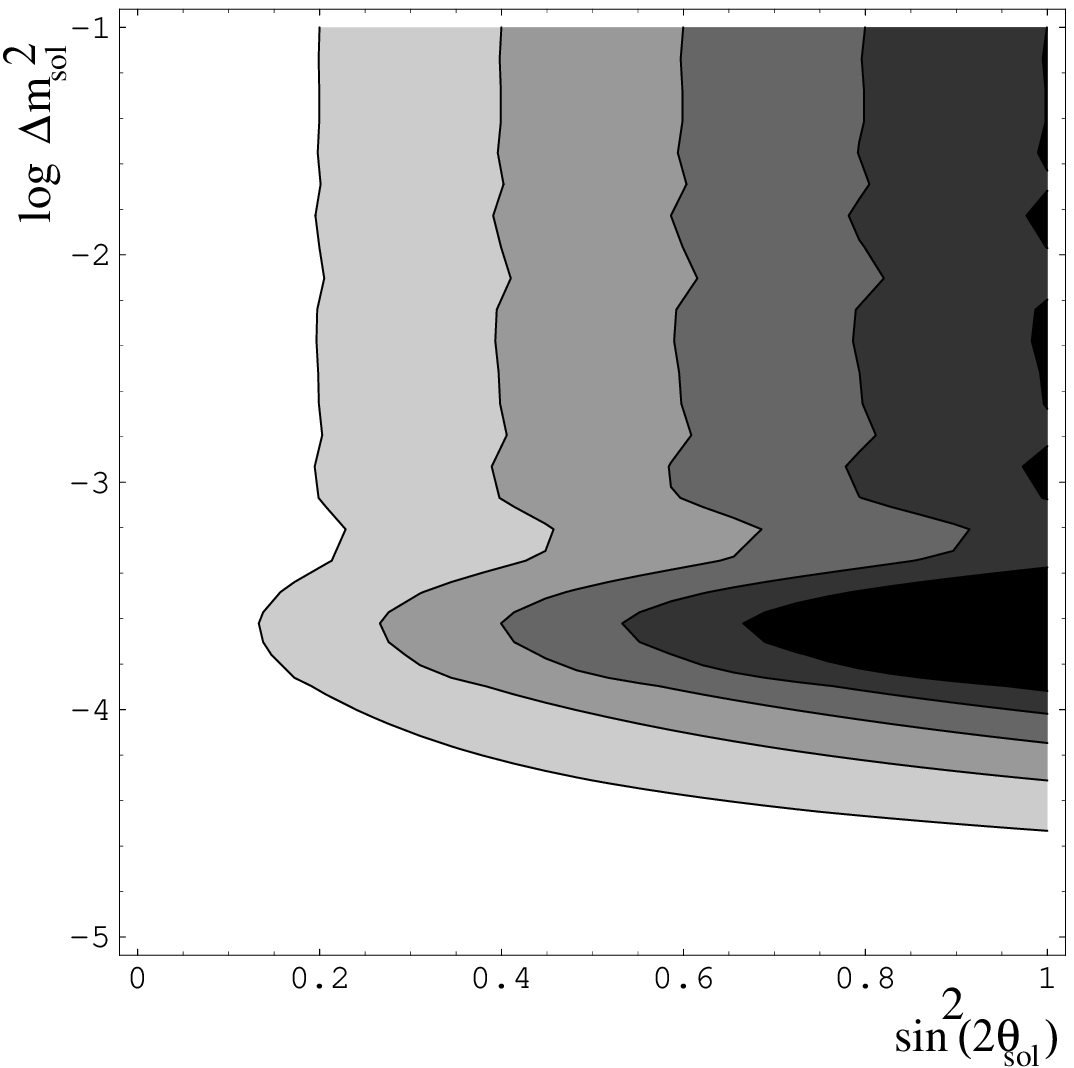, width=5in}{
\label{fig:ratecontour}
{\bf Contour lines of equal suppression by rates only}. The first left curve represents a $10\%$ rate suppression of the $\nuebar$ rate, while each other line marks an additional  $10\%$ suppression step. Assuming no background, and nuclear power plants running at their full capacity, one can detect a $10\%$ rate suppression at a 3~$\sigma$ level after one year of data taking with a detector containing $10^{31}$ target protons.}
\EPSFIGURE[htb]{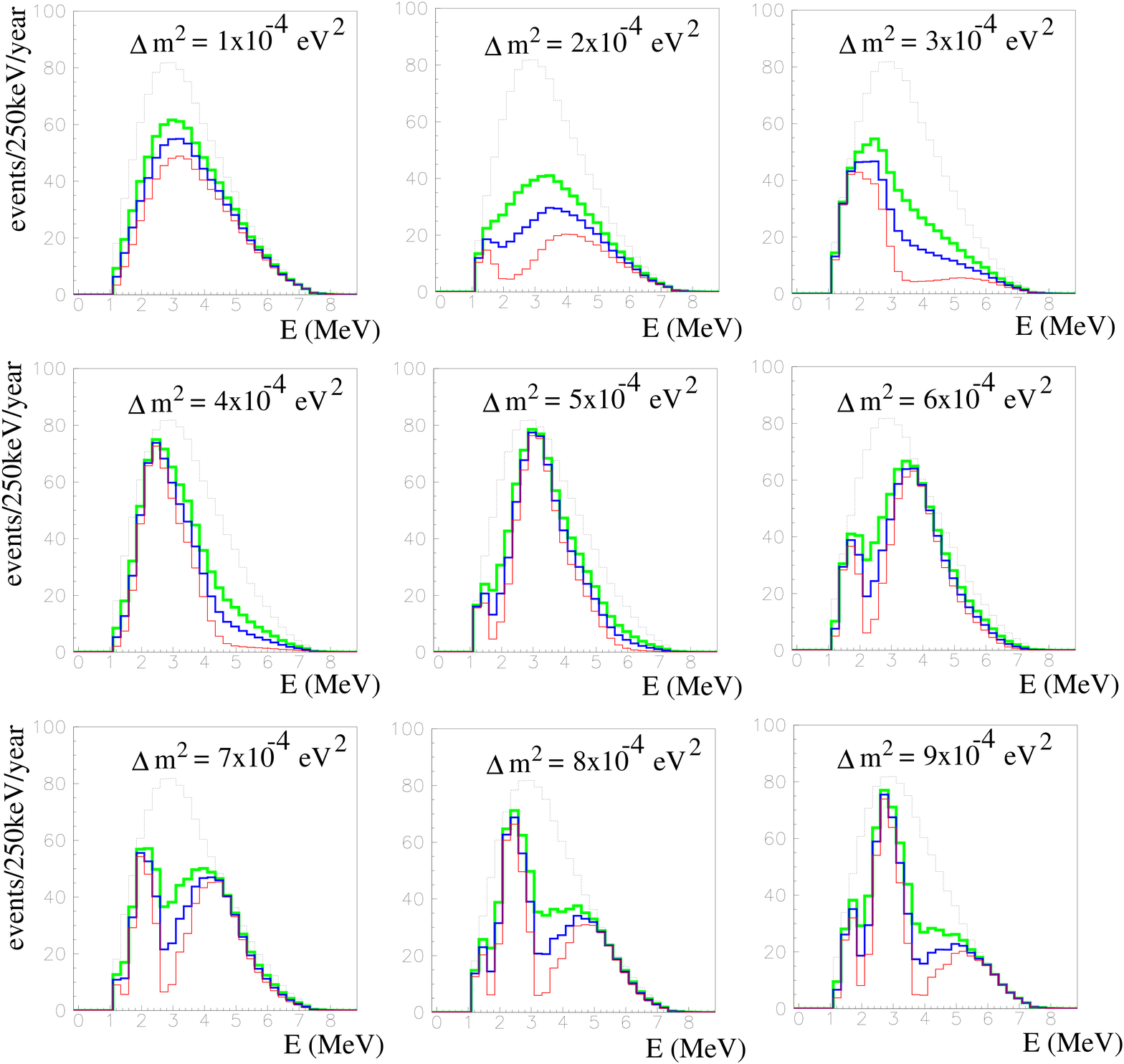, width=6in}{
\label{fig:posspectrum}
{\bf Positron energy spectra for various oscillation solutions} assuming
the power reactors to operate at their nominal power and  a 
detector mass containing $10^{31}$ protons. 
The spectrum is binned in 250~keV intervals. The finite
energy resolution has not been included. The black
dotted line corresponds to the case of no oscillation.
The thick light gray (green) line corresponds $\sin^2 2\Theta_{sol}=0.6$, the medium
thick black line (blue) to  $\sin^2 2\Theta_{sol}=0.8$ and the thin dark gray (red) 
line to $\sin^2 2\Theta_{sol}=1.0$. } 
\subsection{Three-neutrino mixing analysis}

Thanks to its simplicity the two-neutrino oscillation model is a
powerful tool to analyse reactor neutrino data. Nevertheless, on the
baseline of interest ($\sim 20$~km), assuming
$\Dm2_{sol}~\lsim~\Dm2_{atm}$ and $U_{e3} \ne 0$, both $\nuebar$
oscillations due to $\Dm2_{sol}$ (solar driven) and $\Dm2_{atm}$
(atmospheric driven) can develop without being averaged. This would
allow to constrain the relevant $\Ue$ parameter which plays a very important role in the three-neutrino oscillation phenomenology. For instance, it drives the $\numu \leftrightarrow \nue$ ($\numubar \leftrightarrow \nuebar$) oscillations of the atmospheric and (very) long baseline experiments. At present, the most stringent limit comes from the negative result from the CHOOZ experiment\footnote{A slighly less stringent constraint has been obtained by the Palo-Verde experiment \cite{PV}.}; in the case of interest $\Dm2_{sol}~\gsim 2 \cdot 10^{-4}$~eV$^2$, a three-neutrino mixing analysis has constrained $|U_{e3}|^2 < 0.036$ at 95~$\%$ CL, for $\Dm2_{atm} = 2.5 \times 10^{-3}$~eV$^2$ and $\sin^{2}{\Theta_{sol}=0.27}$ \cite{CHOOZU13}. In the forthcoming years, the long baseline accelerator experiments aim to reach a sensitivity for  $|U_{e3}|^{2}$ down to $0.0015$ (see for example \cite{JHF2K}); a somewhat weaker constraint on $|U_{e3}|^{2}=0.003$ is expected from the K2Rdet reactor neutrino experiment \cite{K2RD}. On a much longer timescale, a sensitivity improved by several order of magnitude seems to be achievable with neutrino factories \cite{NuFactories}. Improving our knowledge of  $\Ue$ beforehand is however relevant for the design of neutrino factories. For instance, no {\rm CP}--violation effect would be observable in the lepton sector if $\Ue$ vanishes or is very small. We note here that all these CP--violation effects are unambiguous in long baseline experiment only if $\Dm2_{sol}$ is known with a very high accuracy \cite{Barbieri}. 

In that context, an additional constraint on $\Ue$ could be a by-product of the Heilbronn experiment. So as to illustrate the potential sensitivity to $\Ue$, we compute the positron spectrum obtained at the Heilbronn site with both two and three neutrino  mixing scenarios, for a few $\Ue - \Dm2_{sol}$ combinations. For simplicity, we consider the typical set of LMA parameters $\Dm2_{sol}=5-80\times 10^{-5}$~eV$^{2}$ and $\sin^2{2\Theta_{sol}}=0.8$. Taking into account the SuperKamiokande result on atmospheric neutrinos \cite{SKatm_Messier}, we fix\footnote{ $\Dm2_{atm}$  is assumed to be known at better than $10\%$, and slight modifications of it do not change notably the results.} $\Dm2_{atm} = 2.5 \times 10^{-3}$~eV$^2$. The resulting spectrums are respectively displayed in Fig.~\ref{fig:3nuspectrum_ue3004} and Fig.~\ref{fig:3nuspectrum_ue3002} for $\UeUe = 0.04,0.02$. 
\EPSFIGURE[htb]{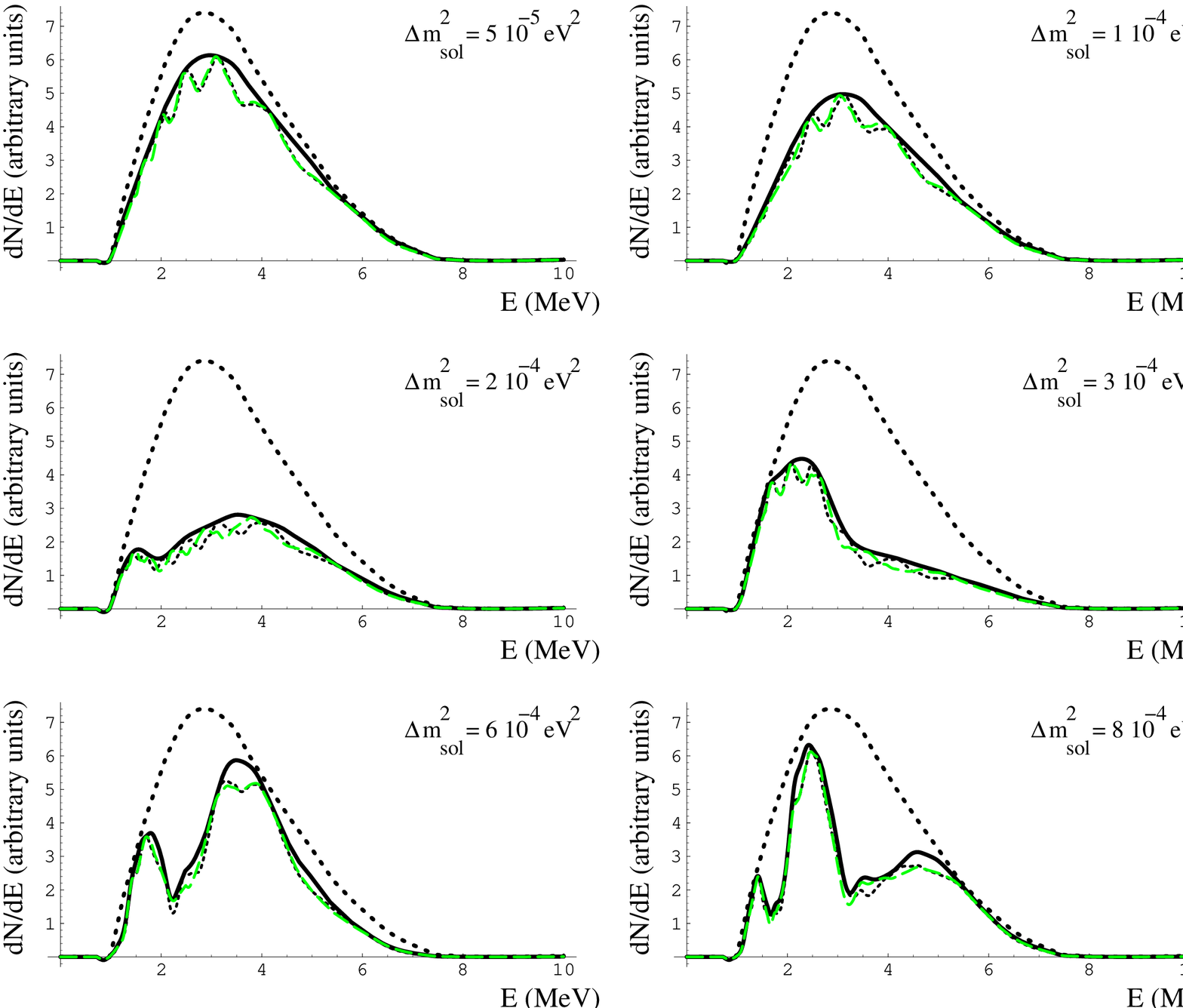, width=5in}{
\label{fig:3nuspectrum_ue3004}
{\bf Difference of the positron spectrum simulated at the Heilbronn
  site, considering the $2\nu$ and $3\nu$ mixing scenarios for various
  oscillation solutions.} The spectrums are not binned and the finite
energy resolution is not included. The thick (black) dotted line
corresponds to the no-oscillation case. The thick (black) line
corresponds to the $2\nu$ oscillation case (the small correction due
to $\Ue$ is included in the effective mixing angle). 
The (black) dotted line and the  thin (green) dashed line represent
the $3\nu$ oscillation case, accounting respectively for the normal
and inverted hierarchy scenario. The parameter choosen are $\sin^2{2\Theta_{sol}}=0.8$, $\Dm2_{atm} = 2.5 \times 10^{-3}$~eV$^2$, and $\UeUe = 0.04$ ($|U_{e1}|^2$ for the inverted hierarchy).}
The envelop of the positron spectrum is roughly given by the two-neutrino solar mixing, whereas ripples are imprinted with the ``atmospheric'' frequency and an amplitude proportional to $\UeUe (1-\UeUe)$ (first and third term of Eq.~\ref{3nuSP}). To detect these ripples one should be able to resolve two adjacent peaks, which are separated by $\Delta(E)$; this requires the energy resolution $\delta {\rm E}$ to satisfy
\begin{equation}
\label{Eres}
\delta  E < \Delta(E) = \left( \frac{a \cdot E_{\nuebar}^2[MeV]}{1 - a \cdot E[MeV]} \right),
%  {\rm with}~ \frac{2\pi}{a} =   2.54 \cdot \Dm2_{atm}[eV^2] \cdot L[m].
\end{equation}
with $2\pi/a =   2.54 \cdot \Dm2_{atm}[eV^2] \cdot L[m]$.
At a baseline of 20 km, for $E_{\nuebar}=3,4,5$~MeV, one has respectively the conditions $\delta E < 0.5,1.0,1.7$ MeV. This seems to be achievable according to the energy resolution quoted by CHOOZ (0.40 MeV), and expected by KamLAND (FWHM  $\sim 2.35 \times 10\% \sqrt{E} \sim 0.50$ MeV) and BOREXINO (FWHM  $\sim 2.35 \times 7\% \sqrt{E} \sim 0.35$ MeV) over the same energy range. In addition, if one does not meet the requirement, the  constraint on energy resolution  could be relaxed by $\sim 25\%$ if one choses the southern site of the Heilbronn mine, located at 14 km of Neckarwestheim (which provides the bulk of the $\nuebar$ flux). 

An adequate statistics is also necessary to detect the ``atmospheric'' driven ripples; to obtain {\it a rough estimate} of the exposure needed, on can subdivide the positron energy spectrum in $\sim 0.5$ MeV energy bins. The statistical error has to be small enough to separate a ``bump'' from an adjacent ``gap'' (i.e 2 bins separated by an amplitude of about $4 \cdot \UeUe (1-\UeUe)$ ), at a $\beta$ sigma level. The expected bin content $n_b$ should then satisfy $n_b \gsim \beta^2/16|U_{e3}|^4$. For $\UeUe=0.04,0.02,0.01$ one obtains respectively  $n_b=40,160,640$ for $\beta = 1$ and $n_b=160,640,2560$ for $\beta = 2$; this leads respectively to   $\sim 0.9,3.5,14 \times 10^{31}$ and $\sim 3.5,14,56 \times 10^{31}$ proton--years of exposure (assuming no background, and power plants running full time at their nominal capacity). Probing $\UeUe$ down to 0.01 corresponds to a $\sim 2\%$ relative effect on average. This is close to the systematic incertainties on the $\nuebar$ energy spectrum production \cite{Bugey}, although those frequencies are expected to be lower than the ``atmospheric'' ripple frequencies. Finally, a complete likelihood analysis will be required to estimate accurately the detection limit of $\Ue$ as a function of   $\Dm2_{sol}$, exposure, and  backgrounds. 

For completeness, it is worthy of mention that the previous discussion remains valid if $\Dm2_{sol} < 2 \times 10^{-4}$~eV$^2$; indeed, if $\Ue$ is not too small, whatever the $\Dm2_{sol}$ value measured at KamLAND, an experiment at the Heilbronn site would observe a global rate suppression from the ``solar'' driven neutrino oscillation and ``atmospheric'' ripples as discussed previously; this is illustrated in the top-left part of Fig.~\ref{fig:3nuspectrum_ue3004} and Fig.~\ref{fig:3nuspectrum_ue3002} for $\Dm2_{sol} = 5 \cdot 10^{-5}$~eV$^2$ and  $\UeUe = 0.04,0.02$.
\EPSFIGURE[htb]{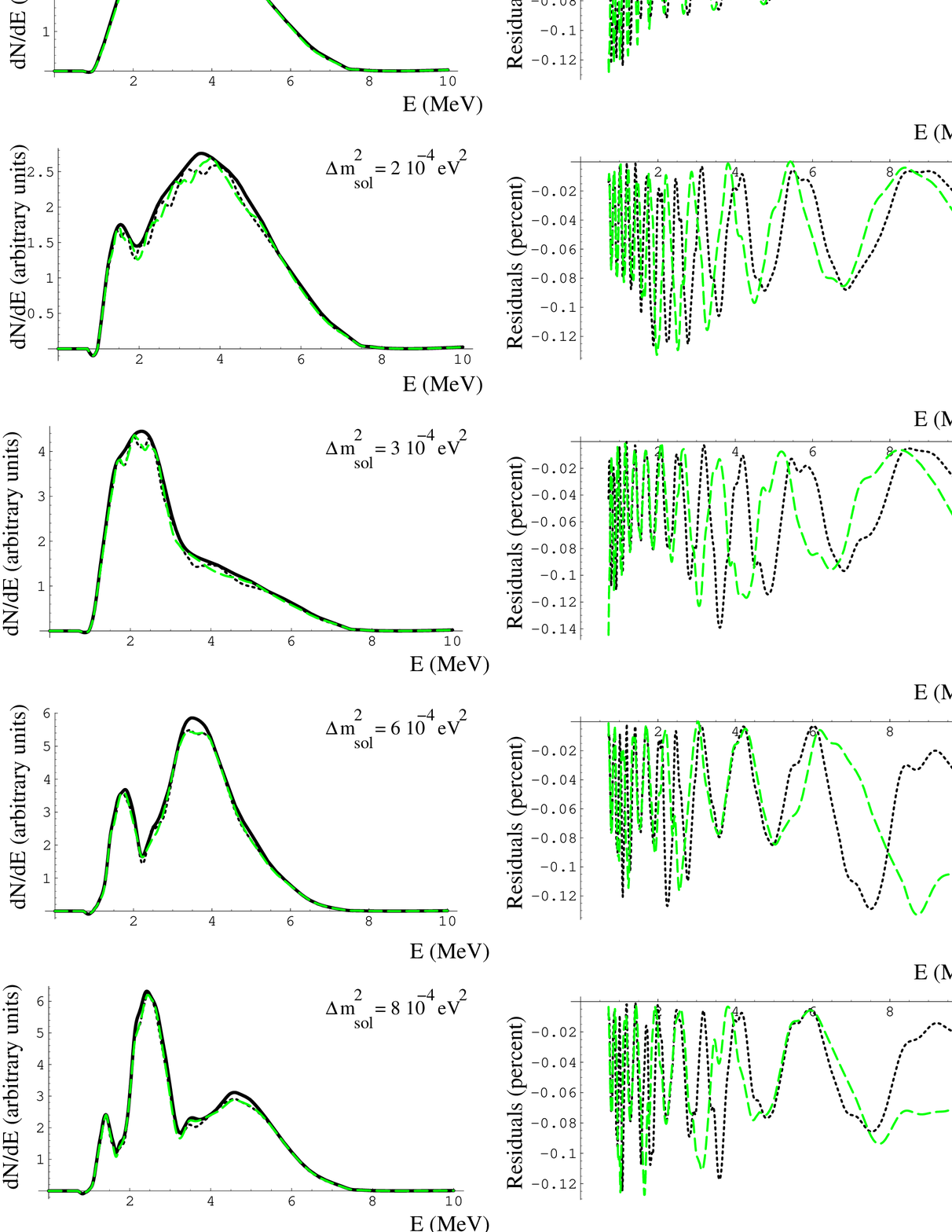, width=5in}{
\label{fig:3nuspectrum_ue3002}
{\bf Three neutrino mixing positron spectrum simulated at the Heilbronn site, for various oscillation solutions (left) and corresponding residuals (Rate(2$\nu$)-Rate(3$\nu$)/Rate(2$\nu$) (right).} The features of these graphs are the same as those of Fig.~\ref{fig:3nuspectrum_ue3004}, except that $\UeUe = 0.02$ ($|U_{e1}|^2$ for the inverted hierarchy).}

As pointed out by Petcov and Piai \cite{PetcovNHIH}, a medium baseline
reactor neutrino experiment could have the potential to distinguish
between normal ($m_1 < m_2 < m_3$) and inverted ($m_3 < m_1 < m_2$)
neutrino mass hierarchy. In this latter case, the heaviest neutrino
field is mainly coupled to the electron neutrino field. To obtain the
survival probability from Eq.~\ref{3nuSP} one has to apply the
following permutations of indices: $2 \rightarrow 3$ and $1
\rightarrow 2$. Eq.~\ref{3nuSP} remains then unchanged (with now
$\Dm2_{32} = \Dm2_{sol}$ and $\Dm2_{atm} = \Dm2_{31}$) apart from the
amplitude of the interference term that can be express as $2
|U_{e1}|^2 (1-|U_{e1}|^2) \cos^2{\Theta_{sol}}$; notice that
$U_{e1}$\footnote{The matrix element that couple the electron neutrino
  to the lightest neutrino field.} is now constrained if one considers
the inverted mass hierarchy model. The difference between the two
hierarchies, for a medium baseline reactor neutrino experiment, is in
principal observable 
if $\sin^2{2\Theta_{sol}} \neq 1$, $|U_{e3/1}|^2 \gsim 0.03$, and
$\Dm2_{atm}$ is known with a high precision \cite{PetcovNHIH}.  At the
Heilbronn site, the net difference between the two kinds of hierarchy
can be seen as a shift of the phase of the ``atmospheric'' driven
ripples on the positron energy spectrum
(Fig.~\ref{fig:3nuspectrum_ue3002}). Since the interference term has
approximately the same frequency as the ``atmospheric'' driven
oscillation term, the energy resolution requirement is also given by
Eq.~\ref{Eres}. 
The exposure needed depends mainly on the $|U_{e3/1}|$ value as discussed  previously. The residuals between the two-neutrino and the three-neutrino cases normalised to the two-neutrino model (Rate(2$\nu$)-Rate(3$\nu$))/Rate(2$\nu$) are shown in the right part of Fig.~\ref{fig:3nuspectrum_ue3002}, for a few $\Dm2_{sol}$ LMA values and $|U_{e3/1}|^2=0.02$. For $\Dm2_{sol}<2 \cdot 10^{-4}$~eV$^2$, the interference term vanishes since $\Dm2_{atm} - \Dm2_{sol} \sim \Dm2_{atm}$ and no difference can be seen between normal and inverted hierarchy. 
%As pointed out by  Petcov and Mai \cite{PetcovNHIH}, the difference between the spectrums in both normal and inverted hierarchy cases is maximal when the solar oscillation survival probability is minimal.% 
Then, Fig.~\ref{fig:3nuspectrum_ue3002} indicates that the difference
between the two mass hierarchies could be distinguished in the narrow
range $\Dm2_{sol}= 2-4 \cdot 10^{-4}$~eV$^2$, while it becomes hardly
detectable for higher $\Dm2_{sol}$ values. 
The ascertainable range is somewhat smaller than the one discussed in 
Ref.~\cite{PetcovNHIH} with $\Dm2_{sol}= 1-5 \cdot 10^{-4}$~eV$^2$. 
Indeed, at the Heilbronn site the interference pattern 
is slightly washed out since reactors at different distances are involved.
\section{Conceptual detector design}

At this stage, we want to line out the basic concepts of a detector
design for the Heilbronn site.
Detector size and energy resolution depend on the physics goals
to be addressed: the determination of $\Dm2_{sol}$ in the HLMA 
parameter range can be performed with a $\sim 100$~ton detector, while
searches for  small oscillation pattern driven by the atmospheric  $\Dm2_{atm}$
with an amplitude proportional to $|U_{e3}|^2$ requires a 
high event statistics, and therefore a large target mass.
We estimate that $\sim 10^{32}$ proton--years are required in 
order to reach a sensitivity of  $|U_{e3}|^2\gsim 0.01$. This implies
a detector similar in size to the KamLAND experiment.
In order to resolve adjacent oscillation peaks, a light yield 
of  $\sim 400$~pe/MeV is required for both detector sizes.
A detailed analysis of the exposure (number of protons~$\times $~years) 
vs. sensitivity in  $|U_{e3}|^2$ needs to be carried out carefully.

The detector concept proposed here is similar to that of   
the Counting Test Facility of BOREXINO \cite{ctfNIM}, however scaled up in size 
and equipped with a hermetical muon veto system.  
It consists of concentrical spherical volumes with a liquid scintillator
target at its center contained in a transparent vessel, 
a water buffer surrounding it, a layer of 
photomultiplier tubes (PMT's) viewing the scintillator target, and optically
separated, an outer water buffer equipped with PMT's as a cosmic ray 
muon detector. A schematic view is displayed in Fig.~\ref{fig:detector_scheme}.

Our current best choice for the liquid scintillator is 
phenyl-xylyl-ethan (PXE, C$_{16}$H$_{18}$) 
as a solvent and p-Tp and bis-MSB as fluor and secondary shifter, or 
alternatively PMP. PXE has a low vapor pressure ($<1.4\times
10^{-4}$~hPa at 20~$^o$C) and a flash point at 149 $^o$C, and
therefore complies with general safety regulations. Moreover, the density is
0.99~g/cm$^3$ and thus creates modest boyant forces to the scintillator
contaiment vessel when emersed in water. 
Comprehensive studies of PXE based scintillators
were performed in the frame of the BOREXINO project comprising
optical properties, radioactive impurities and purification 
methods \cite{BXpxe,rogerPhD,elisaPhD,mariannePhD}. Key results  
include low trace contaminations ($<10^{-17}$g U/g), excellent 
$\alpha - \beta$ discrimination, and high light yield (310 pe/MeV with
20\% optical coverage). 

A $\nuebar$ event is characterized by a prompt positron event 
which deposits a visible energy between 1 and 8 MeV, followed by
a delayed 2.2~MeV gamma event arising from neutron capture in hydrogen
with $\tau \sim 200\, \mu$sec. 
The minimal energy of 1 MeV of the prompt event is due to the positron 
annihilation in the scintillator. Prompt and 
delayed event are spatially correlated (coincidence volume 
$< 1~{\rm m}^3$)
and both have a $\beta / \gamma$-type pulse shape. This characteristic
signature allows to discriminate efficiently against backgrounds.

For the rest of this article, in particular for the estimation 
of the backgrounds, we consider the smallest useful detector size
which still allows high precision determination of $\Dm2_{sol}$ in the 
HLMA parameter range. As generic dimensions we use 300~cm radius 
for the scintillator contaiment vessel, filled with a PXE based scintillator 
(target mass 112~ton), and an optical coverage of about 30~\% providing
a photo electron yield of 400/MeV. Background estimations, as discussed
in the following sections, require the PMT's to be located on a 
sphere with a radius larger than about 500~cm. In order to shield against
external radiation, an additional external water shielding of
about 200~cm is necessary. This can be achieved by a cylindrical tank
of about 14~m in diameter and similar in height. A schematic view
of the detector layout is displayed in Fig.~\ref{fig:detector_scheme}.

As a result of the background considerations we note that a detector
can be realized without using a gadolinium loaded scintillator, since
the accidental background rate can be suppressed  to values well below one
event per year. Therefore, we can achieve the highest possible light
yield with liquid scintillators and hence optimize 
pulse shape discrimination, position reconstruction, as well as
energy resolution.

\EPSFIGURE[htb]{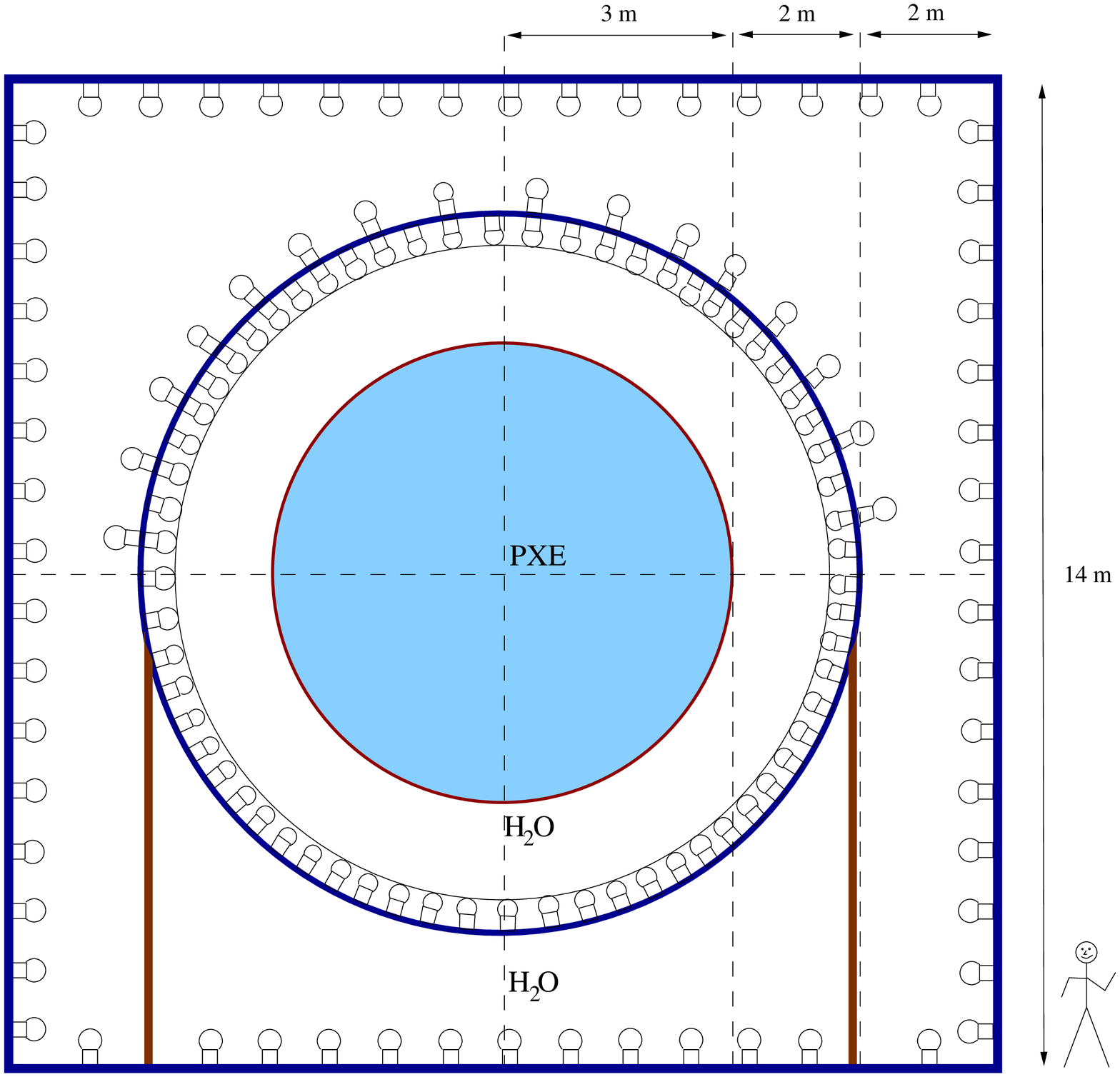, width=6in}{
\label{fig:detector_scheme}
{\bf Schematic view of the proposed detector.}
Depicted is the minimal size of a detector to study the HLMA region
with high accuracy at the Heilbronn underground site. 
About 112 tons of a PXE based
liquid scintillator is contained in a transparent sphere surrounded 
by ultra-pure water as a passive shielding. The design goal is
to achieve a light yield of about 400 pe/MeV which requires an
optical coverage of about 30\% provided by the surrounding PMT's 
and light concentrators. The PMT's are mounted on an open structure
which separates optically the outer part of the detector,
used as a muon veto. At this site, 
the depth varies between 480 and 640 mwe 
reducing the muon flux to about 76 to 36/h/m$^2$.  The $\nuebar$
interaction rate would be about $5\times 10^{2}$ per year
in the no--oscillation case. The detector design allows the 
background rate to be less than one event per year, provided that the 
design specifications of the scintillator and water buffer are met. 
To investigate $\UeUe$ down to $\sim 0.01$ a neutrino target of 
about 1~kt is needed which
implies to enlarge the scintillator sphere to 6.2~m radius. The tank
diameter as well as height would then scale to about 20~m. 
}

\section{Backgrounds}

Oscillation parameters can only be determined with high
accuracy if the backgrounds to reactor neutrino interaction
are small. In particular, a possible observation of small
$|U_{e3}|$ values depends critically on the fraction of 
background events which survive the acceptance cuts.

Backgrounds from primordial and man-made radioactivity can perturb 
neutrino detection, as well as backgrounds induced by cosmic ray interaction. 
Previous developments and experiments provide quantitative guidance
on how to design an experiment which, 
in principle, allows to measure $\nuebar$'s  ``free of background''.
In particular the Counting Test Facility of 
BOREXINO \cite{ctf}, the CHOOZ experiment \cite{CHOOZ2nu} and 
measurements of muon induced production of radio-isotopes \cite{NA54}
provide a quantitative database as well as operational experiences.

\subsection{Non-reactor $\nuebar$ signals}

Further interferences with reactor neutrinos can arise from 
geo-physical $\nuebar$'s.
Electron anti-neutrinos which are emitted in beta disintegration 
of the decay products of 
primordial $^{238}$U and $^{232}$Th in the interior of the earth
have sufficient energy to create background to the reactor neutrino signal.
The visible energy of the geo-$\nuebar$ signal is 
below 2.4~MeV and exhibits a characteristical 
spectral shape. Following Ref.~\cite{geonu}, a rate between 
5 and 33 events per year and $10^{31}$ protons is expected at the Heilbronn 
site{\footnote{Heilbronn is located on the continental
crust. Therefore, the model developed for the interaction rate for 
the BOREXINO detector can be adopted.}. 
The exact value depends on the global abundance of $^{238}$U 
and $^{232}$Th in the earth's crust, mantle and core. First measurements
of the geo-$\nuebar$ flux will be performed by the KamLAND  and 
BOREXINO experiments \cite{geonu,KamPropUS,bx}  and data to estimate the 
interference at the Heilbronn site will be available.
Albeit the geo-$\nuebar$ rate is expected to be 
$\lsim 3\%$ of the reactor rate, 
an interference with the reactor signal at low energies can not be excluded.
A possibility to discard this background is to establish 
an analysis threshold greater than 2.4~MeV.

\subsection{Backgrounds from radioactivity}

Naturally occuring radioactivity can create accidental
as well as correlated backgrounds. 
The radon daughters $^{214}$Bi-$^{214}$Po are a potential 
background of the latter type, because their decay sequence
is similar to the $\nuebar$ tag. 
However, the fast $\beta-\alpha$ decay with $\tau = 237\,\mu$sec can 
be discriminated by the energy deposition of the $^{214}$Po 
$\alpha$-decay, quenched to $\sim 0.8$~MeV (i.e. $>10\, \sigma$ separated
from the delayed 2.2~MeV $\gamma$), 
and in addition by pulse shape information. 

Neutrons from spontaneous fission or ($\alpha,n$)-reactions
can produce recoil protons, subsequently thermalize and
capture on hydrogen atoms. Selection of high purity materials
for detector construction, passive shielding
together with pulse shape discrimination provides an efficient 
handle against this type of correlated background. 
However, high energy neutrons due to cosmic ray muon interaction
may contribute significantly to the background. This is discussed
in detail in Sec.~6.3.2.

Gamma and beta signals in the scintillator volume may generate 
accidentally background events which mimick $\nuebar$ interactions.
The accidental background rate $b_{acc}$ is given by
$b_{acc} \sim b_p b_d \tau_d V_d V_{det}$. 
Here $b_p$ and $b_d$ are the specific background rates
(in units of $sec^{-1} m^{-3}$) for the prompt
and the delayed events, respectively.
The time window for the coincidence is given by $\tau_d$, the
coincidence volume by $V_d$, and $V_{det}$ is the total detetction 
volume.
In order to estimate this background rate in the detector we
use the rather conservative values of $\tau_d \sim 1 \,\, msec$ and
$V_d \sim 1 \,\, m^3$.

If the accidental background rate should be below 
$b_{acc} \sim 1 \,\, y^{-1}$, and hence negligible in comparison
to the neutrino interaction rate,
the condition for the 
background rates for a $V_{det} \sim 113 \,\, m^3$ detector
then reads
$b_p b_d < 3 \cdot 10^{-7} \,\, sec^{-2} \, m^{-6}$.
We are comparing this limit with possible contributions from different 
background sources.

\subsubsection{External backgrounds}

As external background we describe contributions due to the
detector structure material (i.e. PMT's, light concentrators etc.)
and the water shielding outside the liquid scintillator.

The most dangerous source for the external background is the 
strongly penetrating 2.6 MeV gamma line from $^{208}Tl$ in the Thorium chain.
The typical Thorium activity reached in
selected materials for all PMT's and light concentrators
in the solar neutrino experiment BOREXINO is about 
$\sim 10^3 \, Bq$ \cite{bx}.
There a coverage of 30\% at a distance of about 6.5m to the center 
of the detector is achieved. 
Assuming to use PMT's and light concentrators with the same
specific activity 
at a coverage similar to BOREXINO,
the background scales with the surface of the
PM-sphere.

The shielding $S$ of 2.6 MeV gammas
due to the water in a spherical geometry can be parameterized
by $S \sim 10^{-\alpha}$, with 
$\alpha \sim (R-r)/0.5m$, where
$R$ and $r$ are the radii of the PM-sphere and the scintillator 
vessel, respectively.  

Monte-Carlo calculations show that  
signals due to the external 2.6 MeV gamma line contribute 
with about 40~\%  to $b_p$ and to about 10~\% to  $b_d$.
The background rate in the total scintillator
volume can be estimated to be:
$ b_{p-tot} \sim  10^3 \, sec^{-1} \,\, \times (R/6.5m)^2 \times S $. 
For $R \sim 5.0 m$ and $r \sim 3m$ one obtains
$ b_{p-tot} = 2.4 \times 10^{-2} \,\, sec^{-1}$
and $ b_{d-tot} = 6 \times 10^{-3} \,\, sec^{-1}$.
Due to the self-shielding of the scintillator these events are distributed
in an outer region (roughly between 2.5m and 3m) 
with a total volume of about 50$m^3$.
Hence the specific activity at this geometry (i.e. radius of the PM-sphere
at 5m) is $ b_p b_d \sim 6 \cdot 10^{-8} \, sec^{-2} \, m^{-6} $ and therefore 
well below the limit given above.

External backgrounds due to radioactive impurities in the water
shielding have been estimated for Uranium, Thorium, and Potassium
in Monte-Carlo calculations.
For the prompt background all events with energies above 1 MeV 
have been counted.
The energy window for the delayed 2.2 MeV event was chosen
to be 3 sigma at a light yield of 400 photo-electrons per
MeV energy deposition, corresponding to a 30\% optical coverage for
the PMT's and light concentrators.

The critical concentration limits obtained for Th and U 
in the water are in the range
of about $10^{-13} \,\, g/g$ 
to $10^{-12} \,\, g/g$, and for
K between $10^{-10} \,\, g/g$
and $10^{-11} \,\, g/g$, respectively.
The latter obviously gives no contribution to $b_d$, as the
1.46 MeV gammal line from $^{40}K$ is far below the 2.2 MeV
neutron capture energy.

For Thorium the counting rates in the total volume at a concentration of
$5 \cdot 10^{-13} \,\, g/g$ in water are
$b_{p-tot} \sim 1.5 \cdot 10^{-2} \,\, sec^{-1}$ and
$b_{d-tot} \sim 3 \cdot 10^{-3} \,\, sec^{-1}$, respectively.
As these events are distributed mainly in an
outer scintillator volume
of about 20$m^3$, the specific background therefore is
$ b_p b_d \sim  10^{-7} \, sec^{-2} \,\,  m^{-6} $
and below the allowed level given above.

For Uranium the allowed numbers are very similar and one can conclude,
that for both, U and Th, a concentration 
lower than $10^{-12} \,\, g/g$
is sufficient.

These requirements for the water purity have been achieved in the
Counting Test Facility (CTF) of the BOREXINO experiment \cite{ctf}.
Special care has to be taken for Radon in the water, which
is often found to be out of the equilibrium of the U-chain.
The limit given above translates to a tolerable specific Rn-activity
of $\sim 10^2 \, mBq/m^3$.
However, also this (and even better) values have been achieved in the
CTF. 

\subsubsection{Internal backgrounds}

The acceptable single counting rates for U, Th and K
within the liquid scintillator were estimated in
Monte-Carlo calculations.
Also here a 3 sigma window around the 2.2 MeV line for the delayed
event was chosen.
The results show, that the limits for concentrations 
of U and Th are in the level of about $10^{-13} \,\, g/g$.

The results in detail:
at $1 \cdot 10^{-13} \,\, g/g$ for Uranium 
$b_{p-tot} \sim 1 \cdot 10^{-1} \,\, sec^{-1}$ and
$b_{d-tot} \sim 2 \cdot 10^{-2} \,\, sec^{-1}$, 
are the total counting rates in the scintillator volume.
As these events are distributed homogeneously all over
the scintillator volume (i.e. ca. 100$m^3$), the specififc counting 
rates combine to
$ b_p b_d \sim 2 \cdot 10^{-7} \, sec^{-2} \, m^{-6} $,
at the design goal. 
A dominant part of this background comes from $^{214}$Bi decay.
Since the  $^{214}$Bi-$^{214}$Po coincidence can be tagged, 
one can reduce the accidental backgound
further by a about a factor 10. This translates into a 
a limit for the trace contamination of   
$<1 \cdot 10^{-12} \,\, g{\rm U}/g$.

The same limit is basically valid for Thorium. 
As again the K-activity does not contribute to the delayed signal,
the corresponding concentration limits are higher by about 2
orders of magnitude, i.e. in the range of about 
$10^{-11} \,\, g/g$.

In the CTF and with neutron activation analysis
of BOREXINO it has been shown \cite{BXpxe} 
that even better limits have been reached with PXE as liquid solvent 
and p-TP as wavelengthshifter.
The measured values of U and Th of the scintillator as achieved from
the company was already in the $10^{-14} \,\, g/g$ range.
After purification in a Silica-Gel column and using water extraction
limits in Th and U of 
$2 \cdot 10^{-16} \,\, g/g$ and
$1 \cdot 10^{-17} \,\, g/g$
have been obtained, respectively. In Tab.~\ref{tab:Radioactivities}
we summarize the requirements for the principal detector components.

\TABLE[h]{
\centering
\caption{{\bf Summary of background requirements due to 
radioactive trace impurities.} Trace impurities of the PMT's and light 
concentrators have been taken as obtained in the BOREXINO experiment
\cite{bx}. The limit stated for the liquid scintillator impurities
does not include the possibility to discard the   
$^{214}$Bi-$^{214}$Po coincidence on an event by event basis.
}
\label{tab:Radioactivities}
\begin{tabular}{ll}
\noalign{\bigskip}
\hline
Source & Requirements \\
\hline
PMT's and concentrators & distance $\gsim 500$~cm to center \\
Water shield            & $\lsim 5\times 10^{-13}$~g/g, U,Th    \\
                        &  $\lsim 5\times 10^{-13}$~g/g, K      \\
Liquid Scintillator     & $\lsim 10^{-13}$~g/g, U,Th            \\
                        &  $\lsim 10^{-11}$~g/g, K      \\
\noalign{\smallskip}
\hline
\end{tabular}
}

\subsection{Backgrounds induced by cosmic ray muons}

The rock overburden of the Heilbronn site is 180~m at ``Kochendorf'',
the northern area of the Heilbronn saltmine, and 240 m at 
``Neubollinger Hof'', the south western area. 
With a standard rock density of 2.65~g/cm$^3$ the depths correspond
to about 480~m and 640~m of water equivalent (mwe). 
The underground muon 
fluxes, its zenith angle distribution  and mean energies can be calculated 
following Ref.~\cite{bernhard}. The vertical intensity at a depth of 
480 (640)~mwe is $9.8 \,(4.7)\times 10^{-6}\, {\rm cm}^{-2}\, {\rm s}^{-1}\, {\rm sr}^{-1}$ 
and the total flux, taking into account the flat surface topology, is 
$2.1 \,(1.0)\times 10^{-5}\, {\rm cm}^{-2}\, {\rm s}^{-1}$ or 
760 (360)/h/m$^2$.
The mean energy of the vertical flux corresponds to 72 (110)~GeV.

Cosmic ray muons will be the dominating trigger rate at this
depth. About six muons per second will cross a spherical
scintillation target with a volume of 113 m$^3$.
The energy deposition corresponds to about 2 MeV per cm path length
which provides a strong discrimination tool. 

\subsubsection{Muon induced production of radioactive isotopes}
Long-lived muon induced isotopes can not be correlated
to the primary muon interaction if their life time are much longer with
respect to the characteristic time between two subsequent muon interactions. 
They can contribute significantly to the
single trigger rate as well as simulate the 
$\nuebar$ coincidence signature. 

Cross sections of muon induced isotope production on liquid 
scintillator targets ($^{12}$C) have been measured
by the NA54 experiment at the CERN SPS muon beam at
100~GeV and 190~GeV muon energies \cite{NA54}. The energy
dependence was found to scale as 
$\sigma_{tot}(E_\mu)\propto {E_\mu}^{\alpha}$ with 
$\alpha = 0.73 \pm 0.10$ averaged over the various isotopes produced.

Cross sections and interaction rates are calculated for 
a depth of 480~mwe and summarized in Tab.~\ref{tab:Muons}.
\TABLE[h]{
\centering
\caption{{\bf Radioactive isotopes produced by muons and
their secondary shower particles in liquid scintillator targets.}
The rate $R_{\mu}$ is given for a target of 
$5.1\times 10^{30}$ $^{12}$C (with a C/H ratio of 16/18 in 
a pxe-based scintillator this corresponds to a 112 ton target)  at a
depth of 480~mwe. Because of the positron anihilation 
the visible energy in  $\beta^+$ decays is shifted by 1.022~MeV.
$^\ast$: $^9$Li and $^8$He decays could not be separated
experimentally. Therefore, the cross section given, corresponds 
to the sum of   $^9$Li and $^8$He production.}
\label{tab:Muons}
\begin{tabular}{cccccc}
\noalign{\bigskip}
\hline
  &Isotopes & $T_{1/2}$ & $E_{max}$ & $\sigma(72 {\rm GeV})$ & 
$R_{\mu}$ \\
  &         &           & (MeV)     &  ($\mu$barn) &
 (sec$^{-1})$ \\ 
\hline
$\beta^-$ & $^{12}$B & 0.02 s & 13.4 &    n.m.   & - \\
          & $^{11}$Be & 13.80 s & 11.5 &$<0.96$ &  $<8.7\times 10^{-5}$\\
          & $^{11}$Li & 0.09 s & 20.8  &  n.m.   &\\
          & $^{9}$Li  & 0.18 s & 13.6  &$1.0\pm 0.4^{\ast}$&
$(9.2 \pm 3.5) \times 10^{-5}$\\
          & $^{8}$Li  & 0.84 s & 16.0  &$2.3\pm 0.9$ &
$(2.1 \pm 0.8) \times 10^{-4}$\\
          & $^{8}$He  & 0.12 s & 10.6  &$1.0\pm 0.4^{\ast}$ &
$(9.2 \pm 3.5) \times 10^{-5}$\\
          & $^{6}$He  & 0.81 s & 3.5   &$8.0\pm 1.6$ &
$(6.9 \pm 1.7) \times 10^{-4}$\\
$\beta^+$, EC & $^{11}$C & 20.38 min &0.96  &$453\pm 81$ &
$(4.0 \pm 0.8) \times 10^{-2}$\\
          & $^{10}$C  & 19.30 s & 1.9  &$61\pm 10$ &
$(5.5 \pm 0.9) \times 10^{-3}$ \\
          & $^{9}$C   & 0.13 s & 16.0  &$2.4\pm 1.2$ &
$(2.1 \pm 1.1) \times 10^{-4}$\\
          & $^{8}$B   & 0.77 s & 13.7  &$3.3\pm 1.0$ &
$(2.9 \pm 0.9) \times 10^{-4}$\\
          & $^{7}$Be   & 53.3 d & 0.478 (EC, $\gamma$)& $100\pm 20$&
$(9.2 \pm 1.7) \times 10^{-3}$\\ 
\noalign{\smallskip}
\hline
\end{tabular}
}

The dominating background for muon induced 
{\em single} events comes from the decay 
of  $^{11}$C with a rate of about $3.5\times 10^3$ per day or 
$3.6\times 10^{-4}\, {\rm s}^{-1}\, {\rm m}^{-3}$ PXE scintillator.
It thus suffices the constraint on accidental background. 
%The visible
%energy  of the $\beta^+$ decay (including positron anihilation) 
%lies between  1 and 2 MeV. If occuring in coincidence with
%an other single event, it can  either fake a 
%prompt event, or a 2.2 MeV delayed event.  
{\em Correlated} backgrounds induced by muon interactions,
can be created  by $\beta$--neutron instable 
isotopes. Of this type of events, with $^{12}$C as target, 
only $^8$He, $^9$Li and $^{11}$Li  are produced. 
The cumulative production of $^8$He and $^9$Li
together  amounts to about $3\times 10^3$ per year 
for our generic target size. The characteristic signature
of this class of events consists of 
a four-fold coincidence ($\mu \to n \to \beta \to n$). 
The initial muon interaction is followed
by the capture of spallation neutrons within about 1~ms.  The time
scale of the $\beta$-decay of the considered isotopes 
is on the order of a few 100 ms (c.f. Tab.~\ref{tab:Muons}) 
again followed by a neutron capture.  
This signature allows one to discard with high efficiency this class
of events maintaining a high acceptance probability for real
$\nuebar$ events.

\subsubsection{Muon induced neutrons}
Muon induced neutron production can be estimated from the 
results of the CTF experiment \cite{ctf}. At LNGS the muon
mean energy is 320~GeV and the flux corresponds to $1.16 \, {\rm m}^{-2}
\,{\rm h}^{-1}$. 
Scaling the CTF production rate of $0.3\, {\rm d}^{-1}\, {\rm t}^{-1}$ to a 
depth of 480 mwe, we obtain about $7.2\times 10^3 \, {\rm d}^{-1}$. 
About a factor
three higher is estimated following the compilation of 
Ref.~\cite{bernhard}. A muon preceeding a neutron, therefore
must not to be mistaken for a positron event with an efficienecy 
better than one in $10^{6}$ events. 

A further source of background are neutrons which 
are produced in the surrounding rocks by radioactivity and
in cosmic ray muon induced hadronic cascades. 
Neutrons can create a recoil proton and be captured by hydrogen
atoms after thermalization mimicking a $\nuebar$ event.
At depths below several tens of meter, radioactive decays followed by 
$(\alpha,n)$ reactions dominate the overall flux. However,
the energies involved are low and a water buffer of a modest thickness 
efficiently shields these neutrons. High energetic neutrons are produced
in nuclear cascades up to GeV energies and above.
Since the primary cosmic ray muons is not penetrating the 
detector, they are invisible and can not be used for discrimination
purpose. High energetic neutron fluxes have been estimated for the 
Gran Sasso Laboratory and amount to about 25 m$^{-2}$~year$^{-1}$ 
\cite{LVD}. Scaling to the depth of the Heilbronn site 
gives a flux of $5\times 10^{3}$~m$^{-2}$~year$^{-1}$. Assuming a 
surface of 113~m$^2$ 
an overall reduction factor of about $5\times 10^{5}$ is required.
This challenging requirement needs to be obtained by 
passive shielding together with active recognition of the 
recoil nuclei by pulse shape discrimination techniques.

\section{Conclusions}

In this paper we present a possible new 
reactor neutrino experiment with a baseline of $\sim 20$~km,
 dedicated to the accurate determination 
of the solar mixing parameters in the HLMA range 
$  \Dm2_{sol} \gsim  2\times 10^{-4}$~eV$^2$. 
In this parameter range the KamLAND experiment would 
observe only a suppression of the total rate, while
for the proposed experiment the oscillations would become directly
visible as an unique pattern in the positron energy spectrum.
High precision determination of $\Dm2_{sol}$ would then become feasible.
Moreover, this would provide essential input for future long baseline 
neutrino experiments.

We showed that an experiment at the Heilbronn site with a 
minimal detector mass of about 100 tons
could be realized yielding an event rate of about $5\times 10^2$ per year
(in the case of no-oscillation) at a very low background rate of less
than one event per year.
Furthermore, we discussed that small effects related to $\Ue$ could be 
investigated if one increases the detector mass to about 1~kton. 
$\UeUe$  could indeed be constrained down to about 
0.01 and the normal and inverted neutrino mass hierarchies could 
be separated for a distinct combination of parameters.

\bigskip
\acknowledgments
S.S. would like to thank S.~M.~Bilenky for valuable discussions
about three neutrino mixing. We are grateful to  S.~T.~Petcov and M.~Piai
for discussions and for bringing to our attention Ref. \cite{PetcovNHIH}.

\listoftables           % ONLY IN DRAFT MODE
\listoffigures          % ONLY IN DRAFT MODE

\end{document}